\documentclass[10pt, conference]{IEEEtran}
\IEEEoverridecommandlockouts
\usepackage{cite}
\usepackage{amsmath,amssymb,amsfonts}
\usepackage{algorithmic}
\usepackage{graphicx}
\usepackage{textcomp}
\usepackage{xcolor}
\usepackage[linesnumbered,ruled, vlined]{algorithm2e}
\usepackage{subfigure}
\usepackage{booktabs}
\usepackage{amsthm}

\usepackage{geometry}
\usepackage{cite}
\usepackage{amsmath,amssymb,amsfonts}
\usepackage{algorithmic}
\usepackage{graphicx}
\usepackage{textcomp}
\usepackage{xcolor}
\usepackage{tabularx}
\usepackage{multirow}
\usepackage{colortbl}  
\usepackage{xcolor}
\usepackage{booktabs}
\usepackage{amsmath}
\usepackage{marvosym}

\def\BibTeX{{\rm B\kern-.05em{\sc i\kern-.025em b}\kern-.08em
    T\kern-.1667em\lower.7ex\hbox{E}\kern-.125emX}}

\begin{document}

\title{An Attention-Based Denoising Framework for Personality Detection in Social Media Texts
}

\author{\IEEEauthorblockN{Lei Lin}
\IEEEauthorblockA{\textit{Computer Network Information Center} \\
\textit{Chinese Academy of Sciences}\\
Beijing, China\\
linlei@cnic.cn}
\and
\IEEEauthorblockN{Jizhao Zhu*}
\IEEEauthorblockA{\textit{School of Computer Science} \\
\textit{Shenyang Aerospace University}\\
Shenyang, China \\
zhujz@sau.edu.cn
}
\and
\IEEEauthorblockN{Qirui Tang}
\IEEEauthorblockA{\textit{School of Cyber Security} \\
\textit{University of Chinese Academy of Sciences}\\
Beijing, China}
\and
\IEEEauthorblockN{Yihua Du*\thanks{*Corresponding author}}
\IEEEauthorblockA{\textit{Computer Network Information Center} \\
\textit{Chinese Academy of Sciences}\\
Beijing, China \\
yhdu@cashq.ac.cn}
}


\maketitle

\begin{abstract}
In social media networks, users produce a large amount of text content anytime, providing researchers with an invaluable approach to digging for personality-related information. Personality detection based on user-generated text is a method with broad application prospects, such as for constructing user portraits. The presence of significant noise in social media texts hinders personality detection. However, previous studies have not delved deeper into addressing this challenge. Inspired by the scanning reading technique, we propose an attention-based information extraction mechanism (AIEM) for long texts, which is applied to quickly locate valuable pieces of text, and fully integrate beneficial semantic information. Then, we provide a novel attention-based denoising framework (ADF) for personality detection tasks and achieve state-of-the-art performance on two commonly used datasets. Notably, we obtain an average accuracy improvement of 10.2\% on the gold standard Twitter-Myers–Briggs Type Indicator (Twitter-MBTI) dataset. We made our code publicly available on GitHub\footnote{https://github.com/Once2gain/PersonalityDetection}. We shed light on how AIEM works to magnify personality-related signals through a case study.
\end{abstract}


\section{Introduction}
Personality detection is a promising task in both psychology and computer science, with applications in areas such as user portrait construction for targeted marketing and public opinion analysis. User-generated text forms the foundation for automatic personality trait detection. Previous research has demonstrated high within-person stability in language use \cite{schnurr1986methodological,pennebaker1999linguistic,pennebaker2003psychological}, revealing a significant correlation between personality and word use across various text types, including stream-of-consciousness writing, essays, self-narratives, and social media posts \cite{mehl2006personality,hirsh2009personality,ireland2010language,gill2009they,yarkoni2010personality}. Social media blogs and posts, as abundant resources in the digital age, contain rich personality signals, though these are often obscured by noise, presenting both challenges and opportunities for detection.

Early approaches to personality detection focused on analyzing text using ``hand-engineered'' word category dictionaries \cite{pennebaker1999linguistic,pennebaker2001linguistic}, aiming to identify psychologically meaningful word categories that reflect underlying cognitive and emotional processes. With the rise of neural language models, methods like Word2Vec \cite{mikolov2013efficient} and BERT \cite{devlin2018bert} have been employed to learn semantic embeddings for personality detection, achieving promising results \cite{majumder2017deep,mehta2020bottom,jiang2020automatic}. However, a recent study noted that noise in datasets, including insufficient or mixed signals, complicates detection by PLMs\cite{vstajner2021mbti}. Typically, longer texts contain more complete personality information, making it easier for models to identify traits despite noise. However, limitations of input length in BERT and other pre-trained language models (PLMs) restrict their ability to fully leverage longer texts, particularly in social media contexts.

To address this, we adopt a ``divide and rule" strategy, segmenting long texts into manageable parts for sequential processing by PLMs. We then propose a novel attention-based information extraction mechanism (AIEM) that uses shallow text features to identify valuable semantic content and calculate useful combination patterns of segment. Shallow features capture characteristics of the information/semantic content carried by segments without representing the content itself. Leveraging the patterns calculated, we can effectively fuse segment representations generated by PLMs to obtain denoised personality features. Additionally, we incorporate multi-task and ensemble learning strategies into our framework to enhance efficiency and robustness. The main contributions of this study are as follows:

\begin{itemize}
    \item We propose an attention-based information extraction mechanism (AIEM) to overcome the difficulty of finding specific information from long texts. We verify its capacity in a personality detection task to collect personality signals in massive noisy social media texts.
    \item We introduce multi-task learning based on enhanced data to compensate for the lack of training samples and reduce instability. We validate its effectiveness on a small personality detection dataset by comparing it with the AIEM-only method.
    \item We conduct our personality detection experiments by building an ADF and surpassing the current state-of-the-art performance on the Twitter-MBTI dataset by 10.2\%.
\end{itemize}

\section{Related Works}

\subsection{Transfer learning-based method}
Majumder et al.\cite{majumder2017deep} conducted a fundamental study by employing a pre-trained word2vec model for embedding text and a convolutional neural network (CNN) for extracting personality-related text features. After BERT\cite{devlin2018bert} was released, it quickly demonstrated better performance in personality signal learning in texts than traditional methods\cite{keh2019myers,jiang2020automatic,mehta2020bottom}. However, compared to CNN-based personality detection networks, BERT, and such PLMs have limited utilization of user text, which is likely a direction for improvement. Additionally, beyond BERT, traditional machine learning approaches utilizing Extreme Gradient Boosting (XGBoost) with TF-IDF features have shown good results by efficiently capturing psycholinguistic patterns in text\cite{amirhosseini2020machine}. In this paper, we divided the long text, processed each segment separately, and properly integrated signals from different ones to address this gap. Notably, personality detection in texts is closely related to the findings in psycho-linguistics all along\cite{mairesse2007using,poria2013common,mehta2020bottom}. For example, people who scored high on ``extraversion" used more social words, showed more positive emotions, and tended to write more words but fewer large words \cite{pennebaker1999linguistic,tausczik2010psychological}. Linguistic Inquiry and Word Count (LIWC) \cite{pennebaker2001linguistic} is one of the most standard tools for calculating these ``psycho-linguistic features". In transfer learning-based methods, the potential of such features has not been better explored, but rather simply fed into the final classifier of the personality traits along with PLM-based features\cite{majumder2017deep,mehta2020bottom,hans2021text}. In this paper, we used count-based psycho-linguistic features to instruct how to effectively combine semantic features from different text parts to form a comprehensive depict of the author around his thoughts, feelings, and behaviors that profoundly reflect the personality traits.

\subsection{Ensemble learning-based method}
In recent years, ensemble learning has become a widely adopted method for reducing the sensitivity to data noise and improving predictive performance. As for the personality detection task, one way to employ it is to break a long text into multiple segments and vote out the final decision based on the personality predictions over each segment\cite{kazameini2020personality}. The other way is to take predictions based on different types of PLMs into account and make a majority vote\cite{hans2021text,el2022deep,kn2022latent}. The PLMs and ensemble learning-based methods achieved state-of-the-art performance on the two most commonly used datasets, Essays-BigFive and Twitter-MBTI, however, the second method consumes too much computational memory and all the studies do not make specific improvements to the noise problem in personality testing tasks. In this paper, based on the first method’s idea, we designed a denoising framework by integrating multiple approaches, including our uniquely designed AIEM, multi-task learning, ensemble learning, etc., to enhance personality detection performance in long noisy texts. Our framework used a single RoBERTa \cite{liu2019roberta} model and achieved better prediction performance than the state-of-the-art ensemble learning models mentioned above.

\section{Method}

\subsection{Overview}
\begin{figure*}[!h]
    \centering
    \includegraphics[width=0.75\textwidth]{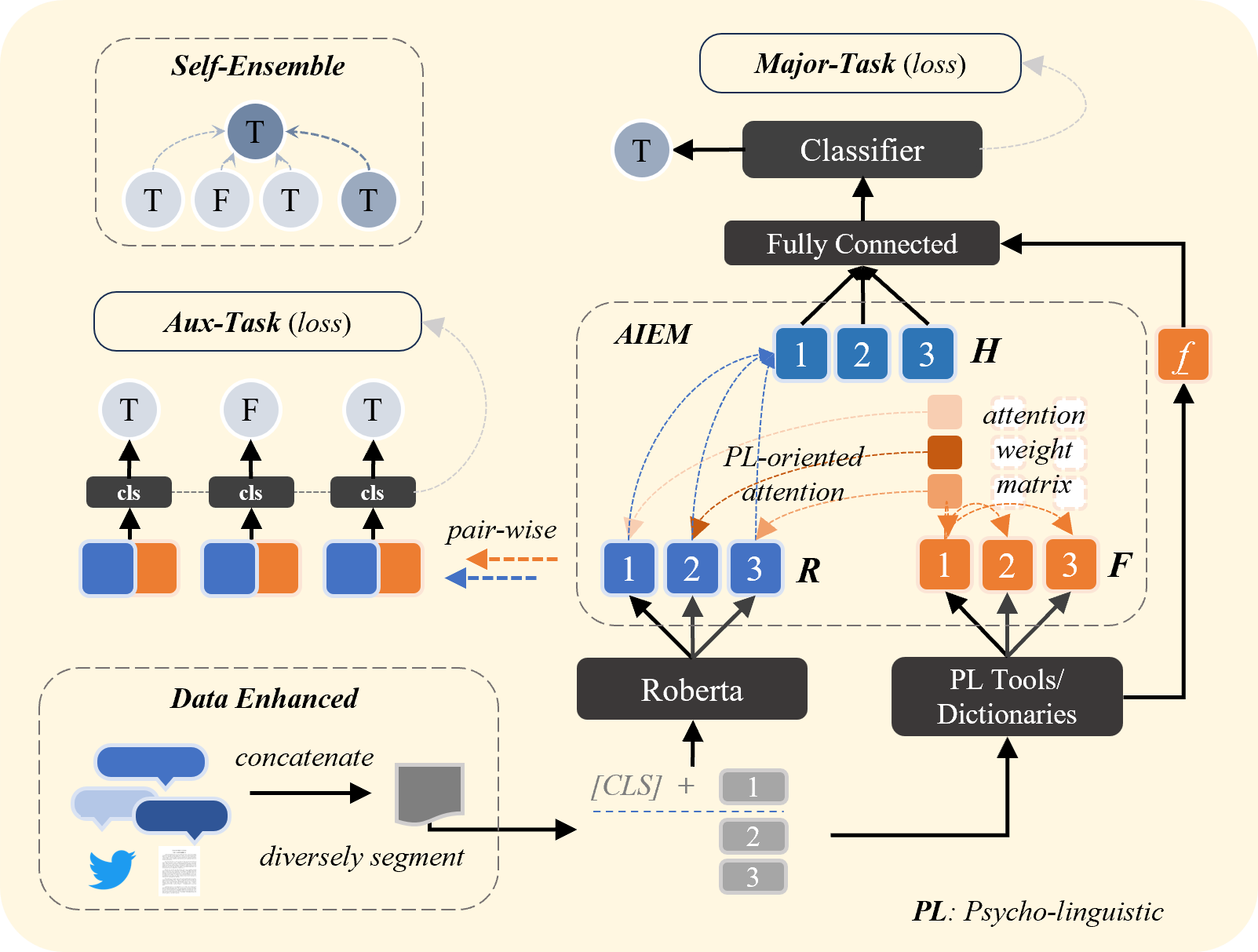}
    \caption{ Sketch of ADF }
    \label{fig:network}
\end{figure*}

We aim to efficiently extract personality-related information from long texts scattered with many unwanted contents. To achieve this, we propose an Attention-based Information Extraction Mechanism (AIEM) that is modified from the attention mechanism \cite{vaswani2017attention}. Due to insufficient personality signals in some samples, personality detection tasks face the challenge of insufficient samples, so we introduce a multi-task learning method to improve the training quality. Besides, using the idea from ensemble learning, we additionally design a self-ensemble method to utilize predictions in aux and major tasks. Overall, we build a denoising personality detection framework, as shown in Figure \ref{fig:network}. 
\subsection{Data enhanced based on segmentation}
\noindent
We prepare our input for AIEM using a data-enhanced strategy, as shown in the bottom ``Data Enhanced" of Figure \ref{fig:network}. Considering that personality is a relatively stable trait whereas noise is sporadic, we uniformly divide a long text sample into multiple segments so that negative signals could be isolated in a small number of segments. For post data, we shuffle all posts of an user, combine them into a complete document-like data, and then perform segmentation.

\subsection{Attention-based information extraction mechanism}
The inspiration for AIEM comes from a speed reading strategy——scanning, which means that you look only for specific pieces of information without reading everything\footnote{https://www.student.unsw.edu.au/reading-strategies}\footnote{https://www.lc.cityu.edu.hk/ELSS/Resource/sas/index.htm}. Scanning involves two key steps summarized below:
\begin{itemize}
    \item moving your eyes quickly down the page to locate specific (key) words, numbers, or data. 
    \item when you locate information requiring attention, you then slow down to read the relevant section more thoroughly. 
\end{itemize}

For the personality detection task, we could seek some ``keywords" to help us locate valuable pieces of text and further integrate relevant semantics to form powerful evidence. In this paper, we deem traditional psycho-linguistic features to be proper ``keywords" as they directly reflect information about emotions, thinking styles, and social concerns, which provide empirical clues on the existence of contents associated with personality manifestation.

\begin{figure}
    \centering
    \includegraphics[width=0.4\textwidth]{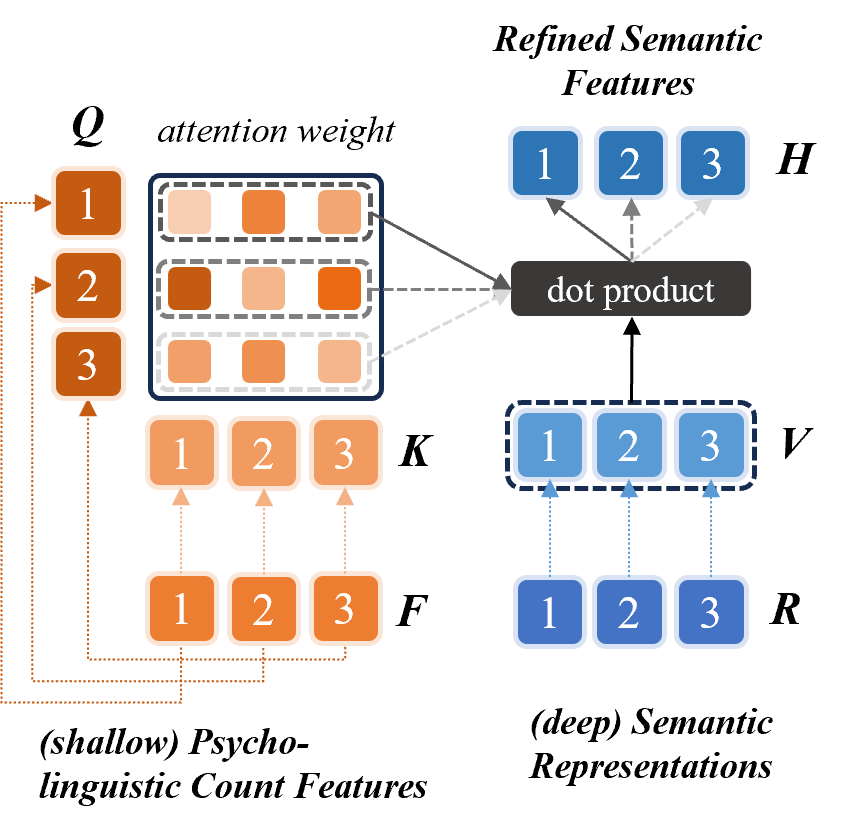}
    \caption{Calculation of  AIEM}
    \label{fig:attention}
\end{figure}

Based on the understanding, we build an AIEM-module over two information vectors shown in the ``AIEM" part of Figure \ref{fig:network}: one is the psycho-linguistic count features vector $F$ and the other is the semantic representations vector $R$. The calculation details of the AIEM are displayed in Figure \ref{fig:attention}. 
Before training, we extract N vectors $f_{1...N} (F)$ of psycho-linguistic count features based on a manually developed ``Counter" program (based on a series of psycholinguistic tools and dictionaries) for each segment $s_{1...N}$ in one sample as well as an overall feature vector $\underline{f}$ for the whole sample $S$ using all texts in it:
\begin{equation} F = Counter(s_{1...N}) ; \quad \underline{f} = Counter(S).
\end{equation}
During training, we use a RoBERTa to learn semantic representations $r_{1...N}(R)$ of $s_{1...N}$:
\begin{equation} R = Roberta(s_{1...N}).
\end{equation}
Notably, we choose the hidden state of the ``[CLS]" token in the last layer as the representation of the input segment. This is a universal choice as ``[CLS]" is added at the beginning of input tokens during pretraining and encodes all information of other words into its hidden state. Based on the attention mechanism \cite{vaswani2017attention}, the attention weight is first calculated as follows: 
\begin{equation} Q,K,V = FW_{q}^ {T},FW_{k}^ {T},RW_{v}^ {T},
\end{equation}
\begin{equation} 
Attn = softmax\left(\frac{{QK^ {T}}}{\sqrt{d_k}}\right).
\end{equation}
$Q$ and $K$ are the query and key required to calculate the attention distribution over the value $V$, respectively. $W_{q}$,$W_{k}$, and $W_{v}$ are the three trainable parameter matrices. $\sqrt{d_k}$ is the dimension of $K$. Finally, semantic information is selectively extracted from $r_{1...N}$ using a weighted fusion of $V$:
\begin{equation} H = Attn\cdot V.
\end{equation}
$H (h_{1...N})$ outputs various combinations of semantic information based on the degree of attention that the segments receive. Similar to scanning, the more a segment is relevant to personality, the more thoroughly it is read.

\subsection{Multi-task learning for personality Detection}
\noindent
To cope with the limited sample condition, we provide an additional learning subject for the network through an auxiliary task (aux task). Specifically, the aux task treats each segment as a training sample and finetunes RoBERTa. The benefits of the aux task are as follows: RoBERTa can quickly adapt to personality detection tasks and generate semantic representation distributions that are conducive to the task.

\subsubsection{Major task}
\noindent
The major task continues fusing features and finishes personality prediction based on the AIEM. Semantic representations $H (h_{1...N})$ extracted by the AIEM module are mapped into shorter vectors using a fully connected (linear) layer. Next, these vectors are concatenated to an overall psycho-linguistic feature vector of the current sample $\underline{f}$ as an advanced feature vector. In our previous study, we established a separate detection model for each trait of the personality model. In this way, the prediction becomes a binary classification question. The end vector is fed into a classifier for a logit value. Then, we adopt the softmax function to generate the probabilities from it. The formulation is described below:
\begin{equation} logit= Classifier\left(Linear(H), \underline{f}\right),
\end{equation}
\begin{equation} P_{major}\left(i\mid x,\sigma\right) = \frac{\exp(logit_i)}{\sum_{j=0}^1 \exp(logit_j)},
\end{equation}
where $i$ is chosen from 0 (negative) and 1 (positive), $\sigma$ denotes the parameters of the model, and $logit_i$ denotes the value of the $i$th position in $logit$. The one in 0 and 1 obtains the highest probability and becomes the label. We adopt Cross Entropy Loss to count the loss of major tasks over input sample $x$ and label $y$:
\begin{equation}\begin{split}
    \mathcal L_{major}\left(x,y\right)
    & = -\sum_{k=1} y_k \cdot \log p_k + (1-y_k) \cdot \log (1-p_k)
    \\& = -\sum_{k=1} \log p\left(y_k\mid H^{\prime}_k, \underline{f_k}, \sigma\right),
\end{split} \end{equation}
where $k$ denotes the index of the sample, $y_k$ denotes the true label of the kth sample, and $p_k$ denotes the probabilities of a positive prediction (refers to 1).

\subsubsection{Auxiliary task}
The aux task supervises the RoBERTa model to predict personality from a single segment. We enable the two classifiers to initialize differently and of different dense layers, mainly for the subsequent self-ensemble method. Each segment's representation vectors $r$ and feature vectors $f$ are fed into either of the classifiers:
\begin{equation} logit_n = Classifier_{0/1}\left(r_n, f_n\right),
\end{equation}
\begin{equation} P_{aux}\left(i\mid s_n,\sigma\right) = \frac{\exp(logit_{n,i})}{\sum_{j=0}^1 \exp(logit_{n,j})},
\end{equation}
where $s_{n}$ denotes the input of the n-th segment in the sample. Similar to the major task, we adopt Cross Entropy Loss to count the loss on each prediction. Finally, the losses are summed across $N$ segments and averaged over $M$ samples:
\begin{equation} \mathcal L_{aux}\left(x,y\right)=-\frac1N\sum_{k=1}^M \sum_{n=1}^N \log p\left(y_k\mid s_{k,n}, \underline{f_{k,n}}, \sigma\right),
\end{equation}
where $k$ denotes the index of the sample and $y_k$ denotes the true label of the kth sample.

\subsection{Self-ensemble of predictions}
Self-ensemble of predictions is an optional approach that plays a role in integrating the predictions of major and aux tasks, as shown in the ``Self-ensemble" part of Figure \ref{fig:network}. Based on previous studies, we expect a more lightweight approach to improve the prediction results, while the two tasks prepare natural inputs for the ensemble. However, there is a slight difference from ensemble learning \cite{sagi2018ensemble} because the ``learner" of each task shares RoBERTa as a basic model. This is why we named our method ``self-ensemble." Self-ensemble is not a completely new concept although the existing understanding of it varies. A few studies in other fields have proposed this method under similar conditions \cite{zhao2019multi,li2022reliability}. Moreover, the aux task is based on a weaker learner compared to a major task and a set of local features. To fix this gap, we changed the way we implement it by narrowing the scope of correcting major predictions. 

According to the aux task, two classifiers each predict half of the segments. We regard the prediction distributions from one classifier as a group. Then, we produce an ensemble prediction $P_{soft,0}$ by soft voting and another $P_{hard,0}$ by hard voting for one group of predictions $P_{half,0}$:
\begin{equation} P_{soft,0} = soft\_vote(P_{half,0}),
\end{equation}
\begin{equation} P_{hard,0} = hard\_vote(P_{half,0}),
\end{equation}
as well as $P_{soft,1}$ and $P_{hard,1}$ for another group of predictions $P_{half,1}$. In hard voting, one of the positive and negative predictions with the highest probability becomes the prediction label, and the most frequently occurring label wins. Soft voting calculates the average probability for each of the positive and negative predictions, and the average one that obtains the highest probability wins.

We correct the major prediction only when the four auxiliary ensemble predictions are the same but opposite to the major prediction:
\begin{equation}
P_{ens} = \left\{ 
\begin{array}{ll}
P_{soft,0}, & \textrm {if $P_{soft,0}$ = $P_{hard,0}$ = } \\
& \textrm {$P_{soft,1}$ = $P_{hard,1}$} \\
& \textrm {and $P_{soft,0}\neq P_{major}$},\\
P_{major}, & \textrm {else}.\\
\end{array} \right.
\end{equation}
Finally, the total loss in the personality detection task can be formulated as follows:
\begin{equation} \mathcal L_{pd}=\lambda_{aux} \mathcal L_{aux} + \lambda_{major} \mathcal L_{major},
\end{equation}
where $\lambda_{aux}$ and $\lambda_{major}$ are weighting factors.

\section{Experiments and Results}

\subsection{Dataset}

We conducted our experiments on two most commonly used publicly available datasets, Essays-Big Five Personality Traits (Essays-BigFive) and Twitter-MBTI dataset. The average length of samples is 672 and 1344, and the maximum length is 2643 and 2014. The descriptions of the two datasets are as follows.

\subsubsection{Essays-BigFive} 
This famous dataset comprises 2468 stream-of-consciousness essays written by students and annotated by self-reports of a 5-factor measure \cite{pennebaker1999linguistic}. It is not directly collected from social media platforms. However, each essay is a ``free writing" small sample, matching most social media text characteristics. This dataset is available on Kaggle\footnote{https://www.kaggle.com/datasets/manjarinandimajumdar/essayscsv} and has been extensively used in personality detection studies. 

\subsubsection{Twitter-MBTI}
This dataset is collected by the Personality Caf'e forum, including 8675 groups of tweets and MBTI labels. Each group comprised the last 50 tweets posted by a user. This dataset is available on Kaggle\footnote{https://www.kaggle.com/datasnaek/mbti-type}. 

\subsection{Psycho-linguistic Features}
Three proven effective psycho-linguistic features in our experiments: Mairesse, SenticNet, and NRC Emotion Lexicon Features\cite{majumder2017deep,mehta2020bottom}. The descriptions of the three features are as follows:

\subsubsection{Mairesse Features}
Mairesse \cite{mairesse2007using} features comprise LIWC \cite{pennebaker2001linguistic} features with 64 dimensions, MRC \cite{coltheart1981mrc} features with 14 dimensions, and prosodic and utterance-type features with 18 dimensions. The LIWC dictionary annotates a word or a prefix with multiple categories involving parts of speech, emotions, society, and environment, while the MRC psycho-linguistic database provides unique annotations in syntactic and psychological information.

\subsubsection{SenticNet Features}
The SenticNet \cite{cambria2022senticnet,susanto2020hourglass}  annotates a word with emotional labels and polarity values. By summing the values of sentiment words that appear in the text, we obtain 5 dimension values corresponding to introspection, temper, attitude, sensitivity, and polarity.

\subsubsection{NRC Emotion Lexicon Features}
The NRC Emotion Lexicon \cite{mohammad2013crowdsourcing} annotates a word with 11 dimension polarity values of anger, anticipation, disgust, fear, joy, negative, positive, sadness, surprise, trust, and charged.

\subsection{Preprocess}

\begin{table}
    \centering
    \caption{Attributes of Processed Datasets}
    \label{tab:processed}
    \begin{tabular}{|c|c|c|c|c|}
    \toprule
    \multirow{2}{*}{Dataset} & \multicolumn{2}{c|}{Num. of Segments} & \multicolumn{2}{c|}{Avg. Word Count} \\\cline{2-5} 
    & small/big & in total & small/big & in total \\ \midrule
    Essays-BigFive & 4/2 & 6 & 170/340 & 227 \\ \midrule
    Twitter-MBTI & 5/3 & 8 & 263/439 & 329 \\ \bottomrule
    \end{tabular}
\end{table}
To meet the different needs of RoBERTa and PL-features Counter, we process datasets in two ways. For RoBERTa, we deeply clean the dataset by removing web links, meaningless numbers or punctuation marks, and other uncommon characters. We try to keep only semantic-related characters because this leads to less training noise. For PL-features Counter, as psycho-linguistic features are based on word (including punctuation marks or other special symbols) frequency count, which is sensitive to all characters written by authors, we keep the nearly raw text to retain all personality-related signals. For each cleaned sample, we divide it into segments of a fixed number. It is worth noting that the segmentation of the Essays-BigFive dataset is more complex as many essays contain unstandardized punctuation marks and are difficult to split into uniformly sized segments. To maintain semantic integrity, we set additional rules, such as using a first-person subject as a split point. To enhance the diversity of information intersection and facilitate the exposure of personality signals, we make random divisions using two sizes. In the end, each sample contains sets of big and small segments, while there may be some overlapping parts between each other. The attributes of datasets after being processed are listed in Table \ref{tab:processed}. 

\subsection{Training Details}

\begin{table}[t]
\centering
\caption{Hyperparameter Settings for Different Personality Detection Datasets}
\label{tab:hyperparameters}
\resizebox{\linewidth}{!}{%
\begin{tabular}{lcc}
\toprule
\textbf{Hyperparameter} & \textbf{Essay-Big Five} & \textbf{Kaggle-MBTI} \\
\midrule
\multicolumn{3}{l}{\textit{Common Settings}} \\
FP16 Precision & True & True \\
Batch Size & 8 & 8 \\
\midrule
\multicolumn{3}{l}{\textit{Training Settings}} \\
Max Epoch & 30 & 15 \\
Max Update & 2,100 & 3,660 \\
Update Frequency & 4 & 4 \\
\midrule
\multicolumn{3}{l}{\textit{Optimizer Settings (Composite)}} \\
\textit{Major-Task Group:} & & \\
Learning Rate & 5e-05 & 5e-05 \\
Weight Decay & 0.1 & 0.1 \\
Warmup Updates & 250 & 300 \\
\textit{Aus-Task Group:} & & \\
Learning Rate & 2e-05 & 4e-05 \\
Weight Decay & 0.15 & 0.15 \\
Warmup Updates & 250 & 300 \\
\textit{Solid Group (for RoBERTa):} & & \\
Learning Rate & 3e-06 & 4e-06 \\
Weight Decay & 0.15 & 0.15 \\
Warmup Updates & 500 & 600 \\
\midrule
\multicolumn{3}{l}{\textit{Model Settings}} \\
Dropout & 0.1 & 0.1 \\
Pooler Dropout & 0.45 & 0.45 \\
\bottomrule
\end{tabular}%
}
\end{table}

We performed three experiments to gradually test the effects of AIEM, the multi-task and self-ensemble method. The models used in the experiments are named ADF, ADF (\&Multi-task), and ADF (\&Self-Ens.) in order, where ``\&" means ``adding a method based on the previous experiment". We used 10-fold cross-validation to verify the generalization ability of the models and trained a model with the best performance for each personality trait, as demonstrated in previous studies\cite{majumder2017deep,hans2021text,el2022deep}. We downloaded the pre-trained RoBERTa-base model from the hugging face and used fairseq \cite{ott2019fairseq} toolkit to build and validate our network. We used a V100 GPU for training. For the multi-task, we observed a fine outcome when setting the value of $\lambda_{major}$ around 1, $\lambda_{aux}$ around 0.1. The comprehensive hyperparameter configurations for both personality detection datasets are detailed in Table~\ref{tab:hyperparameters}. More training details can be found in our repository on GitHub\footnote{https://github.com/Once2gain/PersonalityDetection}.

\begin{table*}[!h]
\centering
\caption{Experiment results on the two datasets}
\label{tab:results}
\begin{tabular}{|l|cccccc|ccccc|}
\hline
\multirow{2}{*}{Model\textbackslash{}Dataset} & \multicolumn{6}{c|}{Essays-BigFive} & \multicolumn{5}{c|}{Twitter-MBTI} \\ \cline{2-12} 
 & EXT & NEU & AGR & CON & \multicolumn{1}{l|}{OPN} & AVE & E/I & S/N & T/F & \multicolumn{1}{l|}{J/P} & AVE \\ \hline
Majority Baseline & 51.7 & 50.0 & 53.1 & 50.8 & \multicolumn{1}{l|}{51.5} & 51.4 & 77.0 & 85.3 & 54.1 & \multicolumn{1}{l|}{60.4} & 69.2 \\ 
Psycholinguistic(PL) Features + MLP & 56.9 & 59.8 & 57.0 & 57.3 & \multicolumn{1}{l|}{60.4} & 58.3 & 77.6 & 86.3 & 72.0 & \multicolumn{1}{l|}{61.9} & 74.5 \\
Bert Embd. + SVM & 57.8 & 58.8 & 57.4 & 56.2 & \multicolumn{1}{l|}{63.2} & 58.7 & 77.0 & 86.2 & 73.7 & \multicolumn{1}{l|}{60.5} & 74.4 \\ \hline
CNN + Mairesse \cite{majumder2017deep} & 58.1 & 59.4 & 56.7 & 57.3 & \multicolumn{1}{l|}{62.7} & 58.8 & / & / & / & \multicolumn{1}{l|}{/} & / \\
Bert Embd. + Mairesse + Bagged-SVM \cite{kazameini2020personality} & 59.3 & 59.4 & 56.5 & 57.8 & \multicolumn{1}{l|}{62.1} & 59.0 & / & / & / & \multicolumn{1}{l|}{/} & / \\ 
XGBoost + TF-IDF Features\cite{amirhosseini2020machine} & / & / & / & / & \multicolumn{1}{l|}{/} & / & 79.0 & 86.0 & 74.2 & \multicolumn{1}{l|}{65.4} & 76.1 \\ 
Bert Embd. + PL Features + MLP\cite{mehta2020bottom} & 60.0 & 60.5 & 58.8 & 59.2 & \multicolumn{1}{l|}{64.6} & 60.6 & 78.8 & 86.3 & 76.1 & \multicolumn{1}{l|}{67.2} & 77.1 \\ \hline
SOTA \cite{el2022deep,kn2022latent} & 61.1 & 62.2 & 60.8 & 59.5 & \multicolumn{1}{l|}{65.6} & 61.9 & 83.4 & 85.6 & 82.2 & \multicolumn{1}{l|}{76.3} & 81.9 \\ \hline
ADF(AIEM) & 61.8 & 61.9 & 60.7 & 60.9 & \multicolumn{1}{l|}{66.4} & 62.3 & 92.1 & 93.6 & \textbf{89.7} & \multicolumn{1}{l|}{91.2} & 91.7 \\
ADF(\&Multi-task) & 63.1 & 63.6 & 61.4 & \textbf{61.2} & \multicolumn{1}{l|}{66.4} & 63.1 & \textbf{92.7} & \textbf{94.7} & \textbf{89.7} & \multicolumn{1}{l|}{91.4} & \textbf{92.1} \\
ADF(\&Self-Ens.) & \textbf{63.6} & \textbf{63.8} & \textbf{61.5} & 61.0 & \multicolumn{1}{l|}{\textbf{66.7}} & \textbf{63.3} & 92.4 & 94.2 & \textbf{89.7} & \multicolumn{1}{l|}{\textbf{91.6}} & 92.0 \\ \hline
\end{tabular}
\end{table*}

\subsection{Overall Performance}

We present the best experimental results in Table \ref{tab:results}. For the Essays-BigFive dataset, ADF achieved an accuracy exceeding 60\% across all five personality traits—a milestone unattained by previous studies.

Models such as CNN + Mairesse\cite{majumder2017deep}, Psycholinguistic features + MLP, and XGBoost + TF-IDF features rely heavily on lexical features. When compared to BERT-based embedding approaches, they obtained comparable results, suggesting that deep semantic features offer similar effectiveness in personality trait prediction.
Notably, the combination of BERT embeddings with psycholinguistic features using MLP improved accuracy from 59\% to 60.6\% on the Essays-BigFive dataset. This indicates that integrating deep semantic and psycholinguistic features can enhance performance in personality recognition. Theoretically, deep semantic features such as BERT embeddings should be more informative in textual understanding. However, in the context of personality recognition, they may also capture semantic noise irrelevant to the task. As shown in Table III, our proposed denoising framework (ADF) significantly boosts classification performance by focusing on text segments that are more discriminative according to psycholinguistic cues.
Furthermore, ADF achieves substantial improvement over the state-of-the-art (SOTA) on the Twitter-MBTI dataset. The longer sample lengths in this dataset allow AIEM to better exploit long-range contextual information, which benefits from ADF’s ability to extract relevant features from extended text inputs.

The results also reveal that the auxiliary task plays a more prominent role in the Essays-BigFive experiments. This aligns with our expectations, as the Essays-BigFive dataset contains fewer samples, limiting RoBERTa’s ability to fine-tune effectively. Additionally, essays are typically written in a single session, lacking the diversity of topics that might help models distinguish between relevant and irrelevant information. As such, RoBERTa requires more data to learn how to separate meaningful from noisy segments.
ADF outperformed previously reported best results in the “EXT”, “CON”, “OPN” traits, as well as in overall average accuracy. Furthermore, ADF(\&Multi-task) and ADF(\&Self-Ensembling) achieved SOTA performance across all five personality traits. For the Twitter-MBTI dataset, ADF surpassed the previous best by approximately 10\%, with ADF(\&Multi-task) and ADF(\&Self-Ensembling) contributing further marginal gains. These models achieved SOTA performance in the “E/I”, “S/N”, “J/P” dimensions and the overall average, while the best accuracy for “T/F” was jointly achieved by all three ADF-based models.

It can be seen that the ADF has made significant progress on the Twitter-MBTI dataset. Twitter MBTI has longer content in samples, which is more conducive for AIEM to leverage the advantages of extracting information from long texts. It can also be observed from the results that the role of the aux task is more significant in the experiments for the Essays-BigFive dataset, which conforms to our understanding. Compared to Twitter-MBTI dataset, Essays-BigFive provides fewer samples for the RoBERTa to adjust the output for the task. Moreover, essays are written in one go, lacking topic changes. RoBERTa needs more samples to learn to widen the representation distance between beneficial segments and unwanted ones.

\subsection{Case Study}

\begin{figure}
    \centering
    \includegraphics[width=0.35\textwidth]{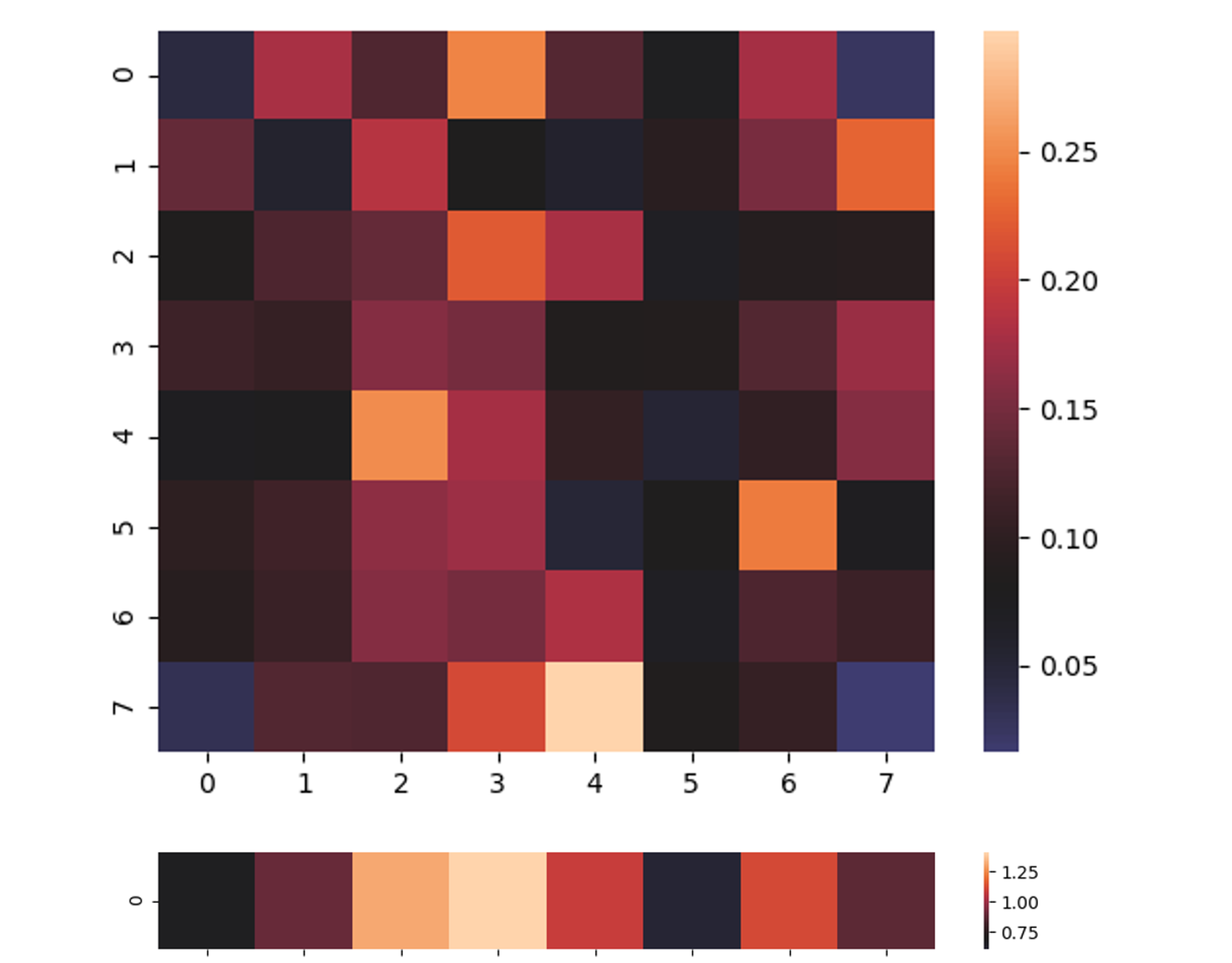}
    \caption{Heatmap of attention weight}
    \label{fig:heatmap}
\end{figure}

To demonstrate how AIEM operates, we pick an ``E" labeled sample from Twitter-MBTI and a trained model to detect EI traits. We feed the sample comprising eight segments into the model and obtained an 8 × 8 attention weight matrix. We transform the matrix into a heatmap, as shown in Figure \ref{fig:heatmap}. 

Each row represents the distribution of attention to all segments by the current segment. It should be noted that the lighter the color, the higher the attention weight obtained. The bottom of the heatmap displays the distribution of the sum of attention weights obtained by each segment.

From the heatmap, we can distinguish the segments with indices 2 and 3 gain the most attention, which means their representations are copied more to the end fused features. From the contents of Segments 2 and 3, we found that they express many open emotions, such as ``\textit{happy}", ``\textit{favorite}", ``\textit{high hopes}", etc, and write many phrases about outdoor activity and social interaction, such as ``\textit{greet hotel staff}", ``\textit{a good nice stroll}", ``\textit{let’s just surf around}", etc. Closely related to this, Segments 2 and 3 contain many semantic statements directly related to the ``extraversion" personality trait. Part of contrent are listed in Table \ref{tab:content}. 

\begin{table}
\centering
\caption{Content in Segments 2 and 3 directly related to the ``extroverted" trait}
\label{tab:content}
\begin{tabular}{ll}
\toprule
Index & Content \\ \midrule
3 & \begin{tabular}[c]{@{}l@{}}Normally when I have lots of free time and nothing\\to do,\textit{I like to go out for a stroll.}\end{tabular} \\ \hline
3 & \begin{tabular}[c]{@{}l@{}}I just happened to wake up 8AM, so \textit{I took a good}\\\textit{nice stroll},  maybe worth of 20km, from 9AM to\\12AM. Made some...\end{tabular} \\ \hline
3 & \begin{tabular}[c]{@{}l@{}}\textit{I love talking about my boobs with female friends},\\we can go with talking our boobs for hours... \end{tabular} \\ \hline
2 & \begin{tabular}[c]{@{}l@{}}\textit{Get a... A totally random man asked me to marry him.} \end{tabular} \\ \hline
3, 2 & \begin{tabular}[c]{@{}l@{}}\textit{So damn happy!} Will be seeing one of my favorite\\bands again next summer!!!!\end{tabular} \\ \hline
3, 2 & \begin{tabular}[c]{@{}l@{}}\textit{I greet hotel staff, cleaners, handymen etc properly}\\\textit{every day} during my stay.\end{tabular} \\ \hline
3, 2 & \begin{tabular}[c]{@{}l@{}}Gonna check out one of local Ju Jutsu dojos\\tomorrow. \textit{I have high hopes, I've heard so much}\\\textit{good stuff about Hokutoryu Ju Jutsu.}\end{tabular} \\ \bottomrule
\end{tabular}%
\end{table}

By contrast, Segments 0 and 5 received the least attention weights. We noticed that there are quite a few emotional words mixed in their texts, but these emotions are difficult to form a unity and what they express is divergent and confusing, such as reflected in sentences ``\textit{When others around you get mad or upset or surprised when you reveal your true feelings and thoughts}" and ``Trying to keep it lively and happy, but I am no longer able to hide my distress, things get ugly". In another aspect, both involve too many discussions about work and sports teams that cannot reveal personality. The complete text of Segments 2,3 and 0,5 can be inquired in Table \ref{tab:weight} in Appendix.

The reasons behind the difference in attention weights lie in the fact that if some segments are unable to provide clear signals that are roughly quantified to psycho-linguistic features later, AIEM is also hard to abstract out a type of personality pattern from these features. As opposed to this, more valuable segments can establish a similarity among each other through AIEM, due to a frequent occurrence of a certain category of emotion or content-related words in texts. 

In other words, a segment retrieval of high attention from others depends on whether its text characters can concretized as a series of public property. That is why noisy segments like Segments 0 and 5 lose its impact on prediction after being processed by AIEM. Although traditional psycho-linguistic features embody a tendency to personality, they are too shallow to be evidence of a certain trait. Fortunately, PLMs detect deep semantics of texts which are selectively magnified by AIEM. 

In conclusion, psycho-linguistic features imply clues associated with personality, while AIEM takes inspiration from these features and produces a possible value distribution over semantic representations of segments. Generally, a piece of text with a set of count-based features that fit into public patterns is more likely to contain decisive semantic content. Benefiting from this, noisy content will be ignored to a certain extent and the negative interference on the result will be reduced.

\section{Conclusion}
\noindent
We propose an Attention-based Information Extraction Mechanism (AIEM) within a broader Denoising Framework (ADF) to improve personality classification from text. By leveraging psycho-linguistic features, multi-task learning, and self-ensembling, our model effectively reduces textual noise. This method significantly outperforms existing benchmarks, improving accuracy by 1.4\% on the Essays dataset and 10.2\% on the Twitter-MBTI corpus.

Despite its advantages, our approach has notable limitations. The current implementation lacks an optimal feature selection mechanism for psycho-linguistic indicators, potentially introducing computational inefficiencies. Additionally, the self-ensemble technique shows inconsistent performance benefits, likely due to the homogeneity of the underlying prediction models. Future work should explore more sophisticated approaches to distinguish personality markers from textual noise, including contrastive learning techniques and domain-specific pre-trained transformer architectures. Investigating the interpretability of attention mechanisms relative to psychological constructs represents another promising direction for interdisciplinary research.

\section{ACKNOWLEDGMENTS}
\noindent
This work was supported by the Strategic Priority Research Program of Chinese Academy of Sciences (XDA0480301), the Informatization Project of Chinese Academy of Sciences (CAS-wx2022gc0304), and the Youth Fund of Computer Network Information Center, Chinese Academy of Sciences (25YF04). We acknowledge the computational resources provided by CAS.

\vspace{40pt}

\appendix
\section{Appendix}
\setcounter{table}{0}
\begin{table*}[!hb]
\centering
\caption{Segments of high and low attention weights}
\label{tab:weight}
\begin{tabular}{|l|p{0.9\linewidth}|}
\hline
Index & Segment Content \\ \hline
3/high & I suspect that my honey is ISFP-ish. Also a person that forces me to lead in relationship. Catastrophe in the making.... So if no one does anything to you, but you just don't like the person, you'd beat him/her up? With no reason at all? Yeah right, you're mental. And your trolling shit is not fucking funny. Get a... A totally random man asked me to marry him. Our customers really xDDD I wonder can one consumpt too much green tea? I've lately addicted on matcha lattes. One cup of matcha is said to equal 10 cups of brewed green tea, and i'm drinking 5-10 matcha lattes daily basis.... Normally when I have lots of free time and nothing to do, I like to go out for a stroll. I just happened to wake up 8AM, so I took a good nice stroll, maybe worth of 20km, from 9AM to 12AM. Made some... So damn happy! Will be seeing one of my favorite bands again next summer!!!!\textless{}3\textless{}3\textless{}3\textless{}3\textless{}3\textless{}3 -Midnight- *let's just surf around a little before going to bed* After a while: *checks clock* 5:30AM Ooooops?:tongue: Gonna check out one of local Ju Jutsu dojos tomorrow. I have high hopes, I've heard so much good stuff about Hokutoryu Ju Jutsu. Why 3 of 5 lightbulbs have to die all at once? Being a shortie (5'0'') changing new ones requires a) ladders or b) tip-toeing on a chair. Still alive after option b. It's not smart way to do it, but... I greet hotel staff, cleaners, handymen etc properly every day during my stay. Everyone treats them as part of the wall, as if they don't exist until you need something. Usually the staff are really... \\ \hline
2/high & The surrounding fields are covered with endless white snow as far as the eye can see. I am very happy. I love winter. :tongue: Everything is snowy white now and the weather is the most brilliant. Loving winter so much. :kitteh: So if no one does anything to you, but you just don't like the person, you'd beat him/her up? With no reason at all? Yeah right, you're mental. And your trolling shit is not fucking funny. Get a... We had it pretty bad for few weeks, but it's quite ok now. Though I live much, much more south-east than you do. \textasciicircum{}\textasciicircum Gonna check out one of local Ju Jutsu dojos tomorrow. I have high hopes, I've heard so much good stuff about Hokutoryu Ju Jutsu. So damn happy! Will be seeing one of my favorite bands again next summer!!!!\textless{}3\textless{}3\textless{}3\textless{}3\textless{}3\textless{}3 I remember you :) -- Mods, if he behaves well from now on, will you show mercy? O.W. Bro Did not know that órkhis means testicles in Greek. Bought an orchis (orkidea) today. Very pretty, I hope I don't kill it accidentally. 9w1 here too :) Whaat... ? I think I'm pretty good holding with bunnies, cats and babies. You just need to make sure that they don't kick, hit or strangle you, and place them on your chest/body, especially if... I'm amused about all these high libido comments. When ya'll turn grumpy and grey old men, and your libido decreases, will you suddenly turn into feminine beings? I know lots of war veterans that... -Midnight- *let's just surf around a little before going to bed* After a while: *checks clock* 5:30AM Ooooops?:tongue: I greet hotel staff, cleaners, handymen etc properly every day during my stay. Everyone treats them as part of the wall, as if they don't exist until you need something. Usually the staff are really... Funny thing is that Baekhyun doesn't bully Tao that much about his skin color although Tao also has same skin tone as Kai. Baekhyun can be extremely stupid and never thinks what he \\ \hline
0/low & I suspect that my honey is ISFP-ish. Also a person that forces me to lead in relationship. Catastrophe in the making.... Kik: domie\_\_\_ Skype: domie\_\_\_ Line: \_\_\_domie WeChat: treeish Kakao: blackmoraltiger (3 underlines) Always welcoming interesting \& random peeps, add away. I need more NTs to my life. . \textless{}--- that's me. When others around you get mad or upset or surprised when you reveal your true feelings and thoughts. Still waters run deep. It's evil witchery, of an introvert dating introvert. Trying to keep it lively and happy, but I am no longer able to hide my distress, things get ugly. Which is really bad considering how badly I... My Fe is butthurt. Hjekjfdfz! I hate when someone is talking back nonsense. Get your shit done and we all could be happy, no? Football \& Introversion Ozil, Ronaldo and football's distrust of introverts | FourFourTwo very good article me thinks. I'd jump into moon for happiness. This would totally work on me. Found a technical bug that no one else had discovered. Hah! Made our IT support speechless, feeling quite proud of myself. Me: *gives negative feedback with normal speech tone* other person: Stop yelling at me! me: to yell means raising one's volume level other person: you're then gifted at nagging quietly me: *sigh*... I wonder can one consumpt too much green tea? I've lately addicted on matcha lattes. One cup of matcha is said to equal 10 cups of brewed green tea, and i'm drinking 5-10 matcha lattes daily basis.... Chewiebon Thanks! Two very lost blueberries haha My best feature though :) Me, me me. Frank Lampard \textless{}3 Second that. I love talking about my boobs with female friends, we can go with talking our boobs for hours... I have some friends who like to touch their friends' boobs, but I personally don't... I'm religious but I don't bother argumenting it or converting anyone. People are just not worth it. Sincerely feel sorry for those who still have faith in people, whether it is for promoting atheism... \\ \hline
5/low & Me: *gives negative feedback with normal speech tone* other person: Stop yelling at me! me: to yell means raising one's volume level other person: you're then gifted at nagging quietly me: *sigh*... Haven't posted here for a while. New piccu from today. Chewiebon Thanks! Two very lost blueberries haha My best feature though :) I work in a small convenience store chain. In our shop there's two regular workers and a supervisor (shop manager). Current shop manager is leaving, she's been talking about it once before, and asked... I'm religious but I don't bother argumenting it or converting anyone. People are just not worth it. Sincerely feel sorry for those who still have faith in people, whether it is for promoting atheism... Wait, what? My internet hubby changed into ENTJ and I only heard about it just now? :D -INTP to the core- Football \& Introversion Ozil, Ronaldo and football's distrust of introverts | FourFourTwo very good article me thinks. When others around you get mad or upset or surprised when you reveal your true feelings and thoughts. Still waters run deep. Sport teams are just like any other corporation, who are dependable for the money their clients bring in. If there were no clients, there will be no business, no profit and hence no team either. That... \\ \hline
\end{tabular}%
\end{table*}

\end{document}